\documentclass[11pt]{article}
\usepackage{moriond,epsfig}
\usepackage{psfrag}

\def\be{\begin{equation}}
\def\ee{\end{equation}}
\def\bea{\begin{eqnarray}}
\def\eea{\end{eqnarray}}
\def\gev{{\rm GeV}}

\def\BXsll{\bar{B} \to X_s l^+l^-}           
\def\be{\begin{equation}}
\def\ee{\end{equation}}
\def\bea{\begin{eqnarray}}
\def\eea{\end{eqnarray}}
\def\nnb{\nonumber}

\newcommand{\BRll}{{{\rm BR}_{\ell \ell}}}

\newcommand{\f}{\frac}
\newcommand{\oas}  {{\cal O}(\alpha_s)}
\newcommand{\scs}{\scriptscriptstyle}

\newcommand{\smallthdm}{\rm{H}}
\newcommand{\s}{\hat{s}}

\newcommand {\ms } {$\overline{\rm{MS}}$ }

\begin{document}
\vspace*{4cm}
\title{$\BXsll$ IN THE STANDARD MODEL \\ AND IN TWO-HIGGS-DOUBLET MODELS}

\author {T. HUBER and S. SCHILLING}

\address{Institute of Theoretical Physics, Winterthurerstrasse 190,\\
8057 Z\"urich, Switzerland}

\maketitle

\vskip -5 cm
\centerline{\includegraphics[width = 3.7 cm]{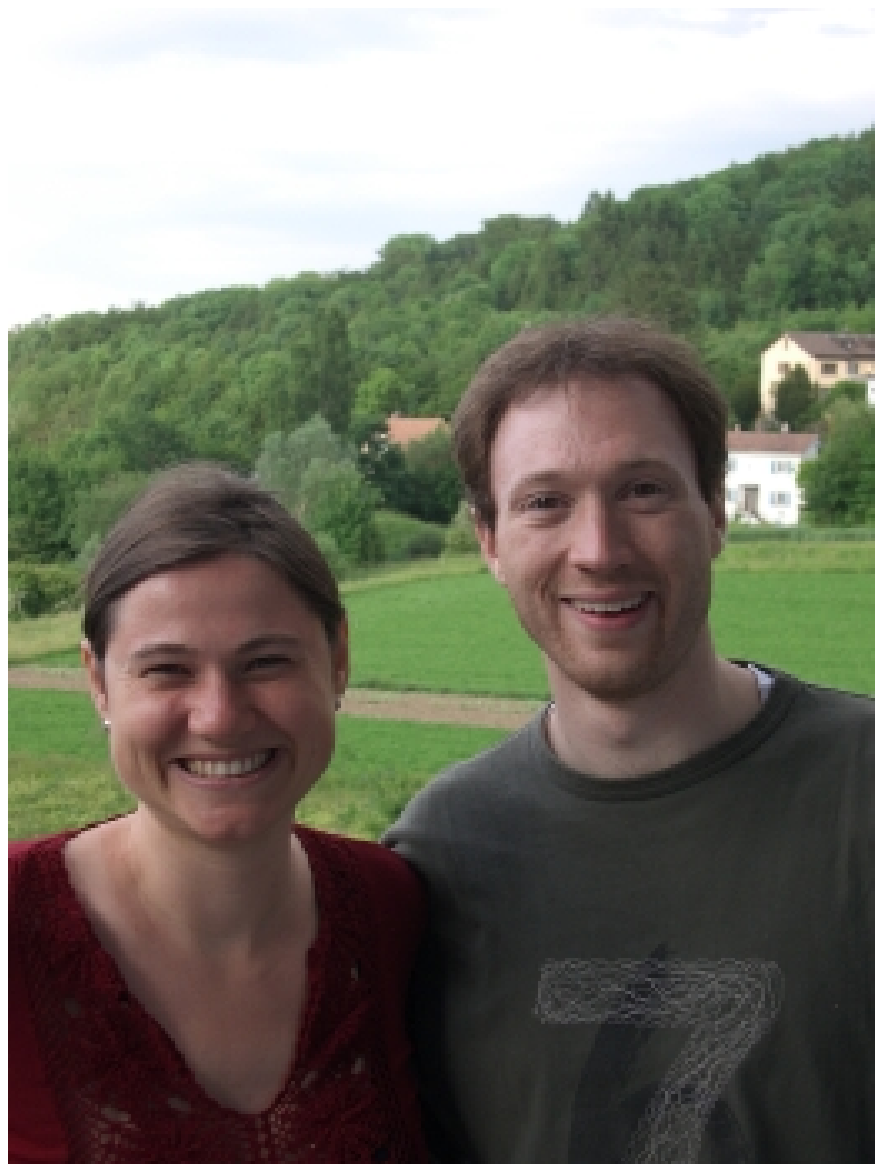} }
\vskip 0.5 cm

\abstracts{
We present recent results of the rare semileptonic decay $\BXsll$. We particularly focus on higher order electroweak corrections in the
Standard Model (SM) as well as $\oas$ corrections in Two-Higgs-doublet models (THDM), both of which are computed within an effective field
theory approach. The calculation of higher order electroweak corrections reveals the presence of enhanced electromagnetic logarithms
$\ln(m_b^2/m_\ell^2)$ in the differential branching ratio. The inclusion of $\oas$ in the THDM reduces the scale dependence of the
corresponding Wilson coefficients significantly.
}

\section{Introduction}\label{sec:introduction}
The inclusive decay $\BXsll$ with $l = e$~or~$\mu$ probes the Standard Model directly at the one loop level and is
therefore sensitive to new physics. Its branching ratio has been recently measured by both
Belle~\cite{Abe:2005.qqqq} and BaBar~\cite{Aubert:2004it}. The experimental results for the differential branching ratio, $\mbox{d}{\cal
B} (\BXsll) \, / \, \mbox{d}q^2$, integrated over the low
dilepton invariant mass region, $1\;\gev^2 < m_{\ell\ell}^2 \equiv q^2 < 6\;\gev^2$, read
\bea
{\cal B} (\BXsll) &=& (1.493 \pm 0.504^{+0.411}_{-0.321})
\times 10^{-6} \;\;\; ({\rm Belle}) \; ,\label{belle}\\ 
{\cal B} (\BXsll) &=& (1.8 \pm 0.7\pm0.5)
\times 10^{-6} \;\;\; ({\rm BaBar}) \; ,\label{babar}
\eea
leading to a world average
\bea
{\cal B} (\BXsll) &=& (1.60 \pm 0.51)\times 10^{-6} \; .\label{average}
\eea
Other appealing features of the decay $\BXsll$ are on the one hand the possibility to obtain complementary information compared to the
less rare decay $\bar{B} \to X_s \gamma$, on the other hand, precision data on both the experimental and theoretical side can be achieved.
Indeed, the experimental errors in the branching ratio are expected to be
substantially reduced in the near future. On the theoretical side, the
predictions are quite well under control because the inclusive hadronic
$\bar{B} \to X_s \ell^+ \ell^-$ decay rate for low dilepton mass is well
approximated by the perturbatively calculable partonic $b \to X_s^{\rm parton}
\ell^+ \ell^-$ decay rate. Thanks to the recent (practically) complete
calculation~\cite{Bobeth:1999mk,Asatryan:2001zw,Asatryan:2002iy,Ghinculov:2003qd,Gambino:2003zm,Gorbahn:2004my,Bobeth:2003at}
of the Next-to-Next-to-Leading Order (NNLO) QCD corrections, the perturbative 
uncertainties are now below 10\%.

However, at the leading order in QED, the branching ratio is proportional to $\alpha_{\mathrm em}^2(\mu)$, giving rise to a $\pm 4$\%
scale uncertainty when the renormalization scale of $\alpha_{\rm em}$ is changed from $\mu = {\cal O}(m_b)$ to $\mu = {\cal O}(M_W)$. 
This uncertainty can be removed by calculating higher order electroweak corrections. The authors of Ref.~\cite{Bobeth:2003at}
calculated the QED corrections to the Wilson coefficients. In a recent calculation~\cite{LunghiMisiakWylerHuber} we confirmed the results
of Ref.~\cite{Bobeth:2003at} and, in addition, computed corrections to the differential branching ratio that originate from QED matrix
elements of four-fermion operators. It turns out that the latter corrections are 
numerically relevant since they are enhanced by large electromagnetic logarithms $\ln(m_b^2/m_\ell^2)$, which originate from these parts
of the QED bremsstrahlung corrections where the photon is emitted collinearly by one of the outgoing leptons.

In the task for the search for new physics beyond the Standard Model it is not only relevant to obtain precise predictions for the
process in question with the Standard Model as the underlying theory, but also to perform precision calculations for extensions of the
Standard Model (SM). In many extensions
of the SM, there are additional one-loop contributions
in which non-SM particles propagate in the loop. If the new particles
are not considerably heavier than those of the SM,
the new contributions to these decays can be as large as
the SM ones. 
One should try to get information on the
parameters in a given extension -- here 
the two-Higgs-doublet models  -- 
from all processes which allow  both a clean theoretical prediction and an 
accurate measurement. This means that 
precision studies similar to those for $B \to X_s \gamma$ 
\cite{Gambino2,BorzumatiGreub,Strumia,MU00},
where higher order QCD corrections are crucial,
should also be done for the process $B \to X_s l^+ l^-$. 
We focus on QCD corrections to the Wilson coefficients $C_9$ and $C_{10}$ in two-Higgs-doublet models (THDM), which
we have calculated in Ref.~\cite{Schilling} and which prove to decrease the scale dependence of the Wilson coefficients
significantly. 
Diagrams with neutral Higgs-boson exchange are neglected.
This omission is justified
in the type-II model, if the coupling parameter
$(m_\ell/M_W) \tan \beta$ is sufficiently smaller than unity. 
In this case the operator basis is the same as in the SM.
Only the matching calculation for the
Wilson coefficients gets changed by adding the contributions where the
flavor transition is mediated by the exchange of the physical charged Higgs
boson. While these extra pieces are known for the
coefficients $C_7$, $C_8$ and $C_{10}$
 to two-loop precision 
\cite{Gambino2,BorzumatiGreub,MU00,Bobeth:2001jm}
for quite some time,
the corresponding results for 
$C_9$ have been first calculated in Ref.~\cite{BobethDis} and were confirmed and first published in Ref.~\cite{Schilling}. 


\section{Theoretical framework}\label{sec:thframework}
\subsection{Effective Theory}
To describe decays like $B \to X_s l^+ l^-$ we use the framework 
of an effective low--energy theory~\cite{buras,buras2,Buras} with five
quarks, obtained by integrating out the heavy degrees of freedom.
In the present case these are the $t$-quark, the $W^{\pm}$ and $Z^0$
boson as well as --- for the THDM calculations --- the charged Higgs bosons $H^{\pm}$, whose masses $M_H$ are
assumed to be of the same order of magnitude as $M_W$.
We only take into account operators up to dimension six and set $m_s=0$.
In  these approximations the effective Lagrangian relevant for our application
(with $ |\Delta B| = |\Delta S| =1$) reads
%
\be \label{Leff}
{\cal L}_{eff} = {\cal L}_{\scs QCD \times QED}(u,d,s,c,b,e,\mu,\tau)  
+\f{4 G_F}{\sqrt{2}} V^*_{ts} V_{tb} 
\left[ \sum_{i=1}^{10} C_i(\mu) {\cal O}_i + \sum_{i=3}^{6} C_{iQ}(\mu) {\cal O}_{iQ} + C_b(\mu) {\cal O}_b \right] .
\ee
The operators ${\cal O}_i(\mu)$ in the first sum are needed for both our 
SM and THDM calculation. They read \footnote{Note that there are several normalizations of ${\cal O}_7$ -- ${\cal O}_{10}$ on the
market.}:
\be
\begin{array}{llll}
{\cal O}_1 \,= &\!
 (\bar{s}_L \gamma_\mu T^a c_L)\, 
 (\bar{c}_L \gamma^\mu T^a b_L)\,, 
               &  \quad 
{\cal O}_2 \,= &\!
 (\bar{s}_L \gamma_\mu c_L)\, 
 (\bar{c}_L \gamma^\mu b_L)\,,   \\[1.002ex]
{\cal O}_3 \,= &\!
 (\bar{s}_L \gamma_\mu b_L) 
 \sum_q
 (\bar{q} \gamma^\mu q)\,, 
               &  \quad 
{\cal O}_4 \,= &\!
 (\bar{s}_L \gamma_\mu T^a b_L) 
 \sum_q
 (\bar{q} \gamma^\mu T^a q)\,,  \\[1.002ex]
{\cal O}_5 \,= &\!
 (\bar{s}_L \gamma_\mu \gamma_\nu \gamma_\rho b_L) 
 \sum_q
 (\bar{q} \gamma^\mu \gamma^\nu \gamma^\rho q)\,, 
               &  \quad 
{\cal O}_6 \,= &\!
 (\bar{s}_L \gamma_\mu \gamma_\nu \gamma_\rho T^a b_L) 
 \sum_q
 (\bar{q} \gamma^\mu \gamma^\nu \gamma^\rho T^a q)\,,  \\[1.002ex]
{\cal O}_7 \, = & \!
 \f{e}{g_s^2} m_b (\bar{s}_L \sigma^{\mu \nu}     b_R) F_{\mu \nu}\,, &\quad
 {\cal O}_8 \, = & \! \f{1}{g_s} m_b (\bar{s}_L \sigma^{\mu \nu} T^a b_R) 
G_{\mu \nu}^a\, ,
 \\[1,00ex]
{\cal O}_9   \,= &\!  \f{e^2}{g_s^2} (\bar{s}_L \gamma_{\mu} b_L) \sum_l 
                                      (\bar{l}\gamma^{\mu} l)\,,
  &  \quad 
{\cal O}_{10}  \,= &\! \f{e^2}{g_s^2} (\bar{s}_L \gamma_{\mu} b_L) \sum_l 
                             (\bar{l} \gamma^{\mu} \gamma_5 l),   

\end{array} 
\label{opbasis}
\ee
where $T^a$ ($a=1,...,8$) are the $SU(3)$ colour generators, and
$g_s$ and $e$ are the strong and electromagnetic coupling constants, respectively. 
$q$ and $l$ appearing in the sums run over the light quarks ($q=u,...,b$) and
the charged leptons, respectively.

Once QED corrections in the SM are considered, five more operators
need to be taken into account. They can be chosen as
\bea
\begin{array}{rl}
{\cal O}_{3Q} = & (\bar{s}_L \gamma_{\mu}     b_L) \sum_q Q_q (\bar{q}\gamma^{\mu}     q),    
\vspace{0.2cm} \\
{\cal O}_{4Q} = & (\bar{s}_L \gamma_{\mu} T^a b_L) \sum_q Q_q (\bar{q}\gamma^{\mu} T^a q),   
\vspace{0.2cm} \\
{\cal O}_{5Q} = & (\bar{s}_L \gamma_{\mu_1}
                     \gamma_{\mu_2}
                     \gamma_{\mu_3}    b_L)\sum_q Q_q (\bar{q} \gamma^{\mu_1} 
                                                               \gamma^{\mu_2}
                                                               \gamma^{\mu_3}     q),
\vspace{0.2cm} \\
{\cal O}_{6Q} = & (\bar{s}_L \gamma_{\mu_1}
                     \gamma_{\mu_2}
                     \gamma_{\mu_3} T^a b_L)\sum_q Q_q (\bar{q} \gamma^{\mu_1} 
                                                                \gamma^{\mu_2}
                                                                \gamma^{\mu_3} T^a q),     
\vspace{0.2cm} \\
{\cal O}_b = & \f{1}{12} \left[ 
          (\bar{s}_L \gamma_{\mu_1}
                     \gamma_{\mu_2}
                     \gamma_{\mu_3}    b_L)            (\bar{b} \gamma^{\mu_1} 
                                                                \gamma^{\mu_2}
                                                                \gamma^{\mu_3}     b)
             -4 (\bar{s}_L \gamma_{\mu} b_L) (\bar{b} \gamma^{\mu} b) \right].
\end{array} 
\eea
where $Q_q$ are the electric charges of the corresponding quarks
($\f{2}{3}$ or $-\f{1}{3}$).

The Wilson coefficients $C_i(\mu)$ are found in the matching procedure by
requiring that conveniently chosen Green's  functions or on-shell
matrix elements are equal when
calculated in the effective theory and in the underlying full theory up
to ${\cal O}[($external momenta and light masses$)^2/M^2]$, where $M$ 
denotes one of the heavy masses like $M_W$ or $ M_H$.
The matching scale $\mu_W$ is usually chosen to be at the order of 
$M$, because at this scale
the matrix elements or Green's functions of the effective operators  pick up the  same large
logarithms as the corresponding quantities in the full theory. 
Consequently, the Wilson coefficients 
$C_i(\mu_W)$ only pick up ``small'' QCD corrections,
which can be calculated in fixed order perturbation theory.

\subsection{Two-Higgs-doublet models}
\label{subsec:THDM}
In the following we consider models with two complex Higgs-doublets 
$\phi_1 $ and $ \phi_2$.
After spontaneous symmetry breaking these two doublets give rise to 
two charged ($H^{\pm}$) and three neutral ($H^0$, $h^0$, $A^0$) 
Higgs-bosons.
When requiring the
absence of flavour changing neutral currents at the tree-level, as we
do in this paper, one obtains two possibilites, the type-I and the type-II
THDM \cite{Weinberg}. 
The part of the Lagrangian relevant for our calculation is
the Yukawa interaction between the charged physical Higgs bosons 
$H^{\pm}$ and the quarks (in its mass eigenstate basis):
\be
{\cal L_I} =\frac{\,g}{\sqrt{2}} \left\{
\left(\frac{{m_d}_i}{M_W}\right)
      X \,{\overline{u}_L}_j V_{ji}  \, {d_R}_i+
\left(\frac{{m_u}_i}{M_W}\right)
      Y \,{\overline{u}_R}_i V_{ij}  \, {d_L}_j
                               \right\} H^+
 +{\rm h.c.}\,.
\label{higgslag}
\ee
The couplings $X$ and $Y$ are 
\bea
\begin{array}{llll}
X=-\cot \beta , & Y &=&\cot \beta \qquad \mbox{(type-I)}, \nonumber \\
X=  \;\;\tan \beta ,& Y &=&\cot \beta \qquad \mbox{(type-II)},
\end{array}
\eea
where $\tan \beta= v_2/v_1$, with $v_1$ and $v_2$ being the vacuum
expectation values of the Higgs doublets $\phi_1$ and $\phi_2$, respectively.


\section{Standard Model corrections to the decay $\BXsll$}\label{sec:SMcorrections}
\subsection{Electromagnetic corrections to the differential branching ratio}\label{sec:emcorrections}

An important quantity in the studies of the rare decay $\BXsll$ is the differential branching ratio, $\mbox{d}{\cal
B} (\BXsll) \, / \, \mbox{d}q^2$, with respect to the invariant mass of the final state lepton pair. The differential branching ratio
as a function of $q^2$ has a region of on-shell intermediate $\bar cc$-resonances like the $J/\Psi$ or the
$\Psi^{\prime}$. This region is therefore not accessible perturbatively and one distinguishes two $q^2$-windows below and above the
$\bar cc$-resonances respectively. The low-$q^2$-window is taken to be from $1\;\gev^2 < q^2 < 6\;\gev^2$, whereas the high-$q^2$-window
is considered for $q^2 > 14.4\;\gev^2$. Many properties --- advantages and disadvantages --- of each window are summarized
in Ref.~\cite{Haisch:2004qs}. We shall restrict ourselves to the low-$q^2$-window here and write the
differential decay width as
\bea
\frac{d\Gamma(\bar B \rightarrow X_s \, l^+ l^-)}{d\hat s} &=& \frac{G_F^2 m_{b, pole}^5 |V^*_{ts} V_{tb}|^2
\alpha_{em}^2(\mu) (1-\hat s)^2}{768 \pi^5} \nnb \\
&&\times \Big\{ \! \! \left(4\!+\!\frac{8}{\hat s}\right) \! \big|\tilde C_7^{eff}\big|^2 \! + (1+2
\hat s) \big(\big|\tilde C_9^{eff}\big|^2  \!\! + \! \big|\tilde C_{10}^{eff}\big|^2  \big) +  12 \, \mbox{Re}\big(\tilde C_7^{eff} \tilde
C_9^{* \, eff}\big) \nnb\\
&&\hspace*{20pt} +   \Delta^{brems}(\hat s)\Big\} \equiv \frac{G_F^2 m_{b, pole}^5 |V^*_{ts} V_{tb}|^2}{48 \pi^3} \cdot
\Phi_{\ell\ell}\, ,\label{diffdecwidth}
\eea 
where we have introduced the notation $\s \equiv q^2/m_b^2$. The effective Wilson coefficients $\tilde C_{i}^{eff}$ contain
all corrections relevant for the calculation up to NNLO in
QCD~\cite{Bobeth:1999mk,Asatryan:2001zw,Ghinculov:2003qd,Gambino:2003zm,Gorbahn:2004my,Bobeth:2003at} and up to NLO in
QED~\cite{Bobeth:2003at,LunghiMisiakWylerHuber}. The last term contains finite gluon bremsstrahlungs corrections~\cite{Asatryan:2002iy}.
Furthermore, we have included the non-perturbative ${\cal O} (1/m_b^2)$ corrections~\cite{Gambino:2001ew,Buchalla:1998mt} and ${\cal O}
(1/m_c^2)$ corrections~\cite{Buchalla:1997ky}.

In order to minimize the uncertainty stemming from $m_{b,{\rm pole}}^5$ and
the CKM angles, we normalize the decay width to the measured semileptonic
one.  Furthermore, to avoid introduction of spurious uncertainties due to the
perturbative $b \to X_c e \bar{\nu}$ phase-space factor, we follow the $\bar
B\to X_s \gamma$ analysis of Ref.~\cite{Gambino:2001ew} where
\be
C = \left| \frac{V_{ub}}{V_{cb}} \right|^2 
       \frac{\Gamma (\bar B\to X_c e\bar\nu)}{\Gamma (\bar B\to X_u e\bar\nu)} 
\ee
was used instead. The factor $C = 0.58 \pm 0.01$ has been recently determined from a global analysis of the semileptonic data
\cite{Bauer:2004ve}. Our expression for the $\bar B \to X_s
\ell^+\ell^-$ branching ratio finally reads
\bea
{{\rm d} {\cal B} (\bar B\to X_s \ell^+\ell^-) \over {\rm d} \hat s} & = &
{\cal B} (B\to X_c e \bar\nu)_{\rm exp} \; 
\left| \frac{V_{ts}^* V_{tb}}{V_{cb}} \right|^2 \; 
\frac{4}{C} \; \frac{\Phi_{\ell\ell}(\s)}{\Phi_u} \;, 
\label{br}
\eea
where $\Phi_u = 1 + {\cal O}(\alpha_s,\alpha_{\mathrm em},\Lambda^2/m_b^2)$ is
defined by
\be \label{bu}
\Gamma (B\to X_u e\bar\nu) =
\frac{G_F^2 m_{b,{\rm pole}}^5}{192 \pi^3} \left| V_{ub}^{}\right|^2 \; \Phi_u.
\ee

As stated earlier, the branching ratio has at leading order in QED a $\pm 4$\%
scale uncertainty due to the renormalization scale dependence of $\alpha_{\rm em}$.
The removal of this uncertainty requires the inclusion of higher order electroweak corrections, namely QED corrections to the Wilson
coefficients~\cite{Bobeth:2003at}. 
These corrections have been known for quite a while and imply that the RGE's for the couplings are coupled differential equations that
have a perturbative expansion in $\alpha_s$ and $\alpha_{\rm em}$. In Ref.~\cite{LunghiMisiakWylerHuber} the results of
Ref.~\cite{Bobeth:2003at} for all the two-loop anomalous dimension matrices that are relevant for the running of the Wilson coefficients
from high scales of order ${\cal O}(\mu_W)$ down to scales of order ${\cal O}(\mu_b)$ were confirmed.

In addition, corrections to the differential branching ratio that originate from QED matrix
elements of four-fermion operators were computed~\cite{LunghiMisiakWylerHuber}. The loop corrections are not free of infrared divergences
and must therefore be considered together with the corresponding bremsstrahlung. The dilepton invariant mass differential decay width is
not an infrared safe object with respect to emission of collinear photons. Hence,
QED corrections contain an explicit electromagnetic logarithm $\ln(m_b^2/m_\ell^2)$, which stems from these parts of the QED
bremsstrahlung corrections where the photon is emitted collinearly by either of the final state leptons. These enhanced terms always arise
as single electromagnetic logarithms accompanied by an electromagnetic coupling $\alpha_{\rm em}$ and therefore do not get resummed.
The log-enhanced parts of the QED matrix elements disappear after integration over the whole phase space available but survive and remain
numerically important when we restrict $q^2$ to the low dilepton invariant mass region, $1\;\gev^2 < q^2 < 6\;\gev^2$, that we consider.
Their numerical impact on the differential branching ratio integrated over the low-$q^2$ window is about
+5 \% for final state electrons and about +2 \% for final state muons. The numerical results for the branching ratio integrated over the
low-$q^2$ region read
\bea
\hskip -0.5cm
{\cal B}_{\mu\mu} & = & \label{muonBR} 
( 1.59 \pm 0.11) \times 10^{-6} \;, \\
\hskip -0.5cm
{\cal B}_{ee} & = & \label{electronBR} 
(  1.64 \pm 0.11) \times 10^{-6} \;.
\eea
However, the large effect for electrons gets reduced in size
roughly to that of muons once the experimental resolution for collinear photons is taken into account. Precise numbers and more profound
explanations on the advent of the collinear logarithm are given in Ref.~\cite{LunghiMisiakWylerHuber}.

\subsection{The forward backward asymmetry ${\cal A}_{FB}$}\label{sec:AFB}

Another appealing quantity of the decay $\BXsll$ is the so-called forward backward asymmetry ${\cal A}_{FB}$ defined as
\be\label{AFB}
{\cal A}_{FB}(q^2) \! \equiv \frac{d\BRll/d q^2(\cos\theta_\ell>0) \, -  \, d\BRll/d
q^2(\cos\theta_\ell<0)}{d\BRll/d q^2(\cos\theta_\ell>0) \,  + \, d\BRll/d q^2(\cos\theta_\ell<0)} \; ,
\ee
where $\theta_\ell$ is the angle between the positively charged final state lepton and the initial state $B$-meson in the restframe of the
final state lepton pair. The forward backward asymmetry is also nicely reviewed in Ref.~\cite{Haisch:2004qs}. As it is defined as a
difference of two quantities over the corresponding sum, it is almost insensitive to hadronic uncertainties since the latter tend to
cancel in the ratio. In the SM, the forward backward asymmetry has a zero at~\cite{Bobeth:2003at,Haisch:2004qs}
\be\label{q02}
q_0^2 = (3.76 \pm 0.22_{\rm theory} \pm 0.24_{m_b})\, \gev^2
\ee
which is also subject to receive contributions from large electromagnetic
logarithms~\cite{HuberAFB}.

The branching ratio and the forward backward asymmetry of the decay $\BXsll$ are important for yet another reason. Contrary to the
branching ratio of the rare decay $\bar{B} \to X_s \gamma$, which is at lowest order proportional to $\big|\tilde
C_7^{eff}\big|^2$, both the branching ratio of $\BXsll$ and the forward backward asymmetry are sensitive to the sign of $\tilde
C_7^{eff}$. Changing the sign of $\tilde C_7^{eff}$ results on the one hand in a shift of the branching ratio. This shift is so large
that the value of the branching ratio gets moved out of the experimentally allowed range. Therefore the SM sign of $\tilde
C_7^{eff}$ is favored~\cite{Gambino:2004mv}. On the other hand, a change of the sign of $\tilde C_7^{eff}$ removes the
presence of a zero in the forward backward asymmetry~\cite{Ali:2002jg}. Hence already a rough measurement of the branching ratio and the
shape of the forward backward asymmetry can yield useful information about the sign of $\tilde C_7^{eff}$. The determination of the sign of
$\tilde C_7^{eff}$ is crucial since it allows to strongly constrain the parameter space of certain SUSY
models~\cite{Gambino:2004mv,Cho:1996we}.


\section{QCD corrections to the Wilson coefficients $C_9$ and $C_{10}$ in the THDM}
\label{sec:THDM}
\begin{figure}
\begin{center}
\epsfig{file=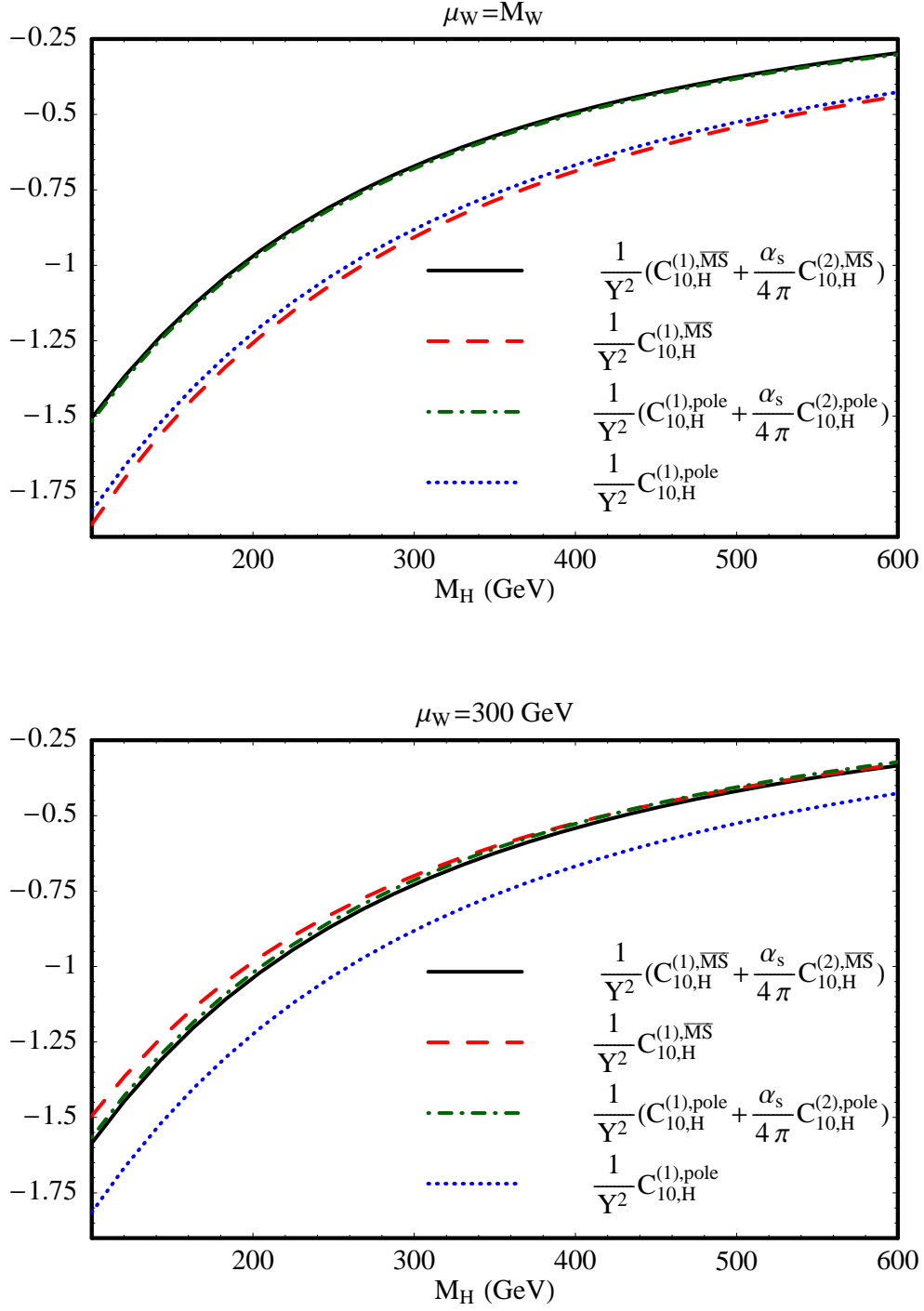,width=15.5cm}
\end{center}
\caption{Dependence of the rescaled Wilson coefficient 
$\hat{C}_{10,\smallthdm}(\mu_W)$ 
(see Eq. (\ref{rescaled})) on the charged Higgs boson mass $M_H$
at the matching scale
$\mu_W=M_W$ (upper frame) and $\mu_W=300$ GeV (lower frame).
The dashed (dotted) line is the one-loop contribution expressed in
\ms -scheme (pole-mass scheme) of the $t$-quark mass, 
while the solid (dash-dotted) line includes the two-loop corrections  in the
respective scheme.}
\label{combi}
\end{figure}

For the following it is
convenient to expand the Wilson coefficients $C_i(\mu_W)$ as 
\be 
\label{expanded.coeffs}
C_i(\mu_W) = C^{(0)}_i(\mu_W) 
+ \f{\alpha_s}{(4 \pi)} C^{(1)}_i(\mu_W)  
+ \f{\alpha_s^2}{(4 \pi)^2} C^{(2)}_i (\mu_W) 
+ {\cal O}(g_s^6).
\ee
The analytic formulae for the 
QCD corrections to the Wilson coefficients $C_9$ and $C_{10}$ in the THDM
at the matching scale $\mu_W$ are given in Ref.~\cite{Schilling}. 
In this section we briefly illustrate the impact of these two-loop corrections on $C_{10,\smallthdm}(\mu)$.
We then introduce a rescaled Wilson coefficient (see Eq. (\ref{expanded.coeffs}))
\begin{equation}
\label{rescaled}
\hat{C}_{10,\smallthdm}(\mu_W) \doteq \frac{1}{Y^2} \,
\frac{4\pi}{\alpha_s(\mu_W)} 
C_{10,\smallthdm}(\mu_W).
\end{equation}
In Fig. \ref{combi}
we plot the quantities
\be
\frac{1}{Y^2} C_{10,\smallthdm}^{(1)}(\mu_W) \quad \mbox{and} \quad
\frac{1}{Y^2} \left( C_{10,\smallthdm}^{(1)}(\mu_W) + 
\frac{\alpha_s(\mu_W)}{4\pi}
                     C_{10,\smallthdm}^{(2)}(\mu_W) \right)  ,
\ee
i.e. two approximations of $\hat{C}_{10,\smallthdm}$
as a function of the charged Higgs boson mass $M_H$
for the $\overline{\rm{MS}}$- and for the pole mass scheme of the $t$-quark
mass. As input parameters we use 
$\alpha_s(M_Z) = 0.119$,
$m_t^{\rm{pole}} = 178.0~{\rm GeV}$,   
$M_W = 80.4~{\rm GeV}$ and         
$s^2_W = 0.231$ \cite{PData02,nature}.
The upper frame shows these quantities at the relatively low matching scale
$\mu_W=M_W$. As in this case $m_t^{\rm{pole}}$ and $\overline{m}_t(\mu_W)$ 
are numerically
almost identical, the one-loop approximations (dotted and dashed lines) 
are close
to each other. The inclusion of the two-loop corrections, however, 
considerably lowers
the (absolute) size of the coefficient for all values of $M_H$ considered. 
In the lower
frame a higher matching scale of $\mu_W=300$ GeV is chosen. As in this case
$m_t^{\rm{pole}}$ and $\overline{m}_t(\mu_W)$ differ considerably, 
the renormalization
scheme dependence of the one-loop results is rather large. When taking into
account the two-loop corrections (solid and dash-dotted lines), 
the scheme dependence is drastically reduced.

\begin{figure}
\psfrag{legend1}
{{$\frac{1}{Y^2}\left(C_{10,H}^{1,\overline{MS}}+
 \frac{\alpha_s}{4 \pi}  C_{10,H}^{2,\overline{MS}}\right)$}}
\psfrag{mh300}{{$M_H=300\; GeV$} }
\psfrag{mh600}{{$M_H=600\; GeV$} }
\psfrag{muw}{{$\mu_{W} \;(GeV)$}}
\psfrag{legend2}{{ $\frac{1}{Y^2}\;\;\;C_{10}^{1,\overline{MS}}$}}
\begin{center}
\epsfig{file=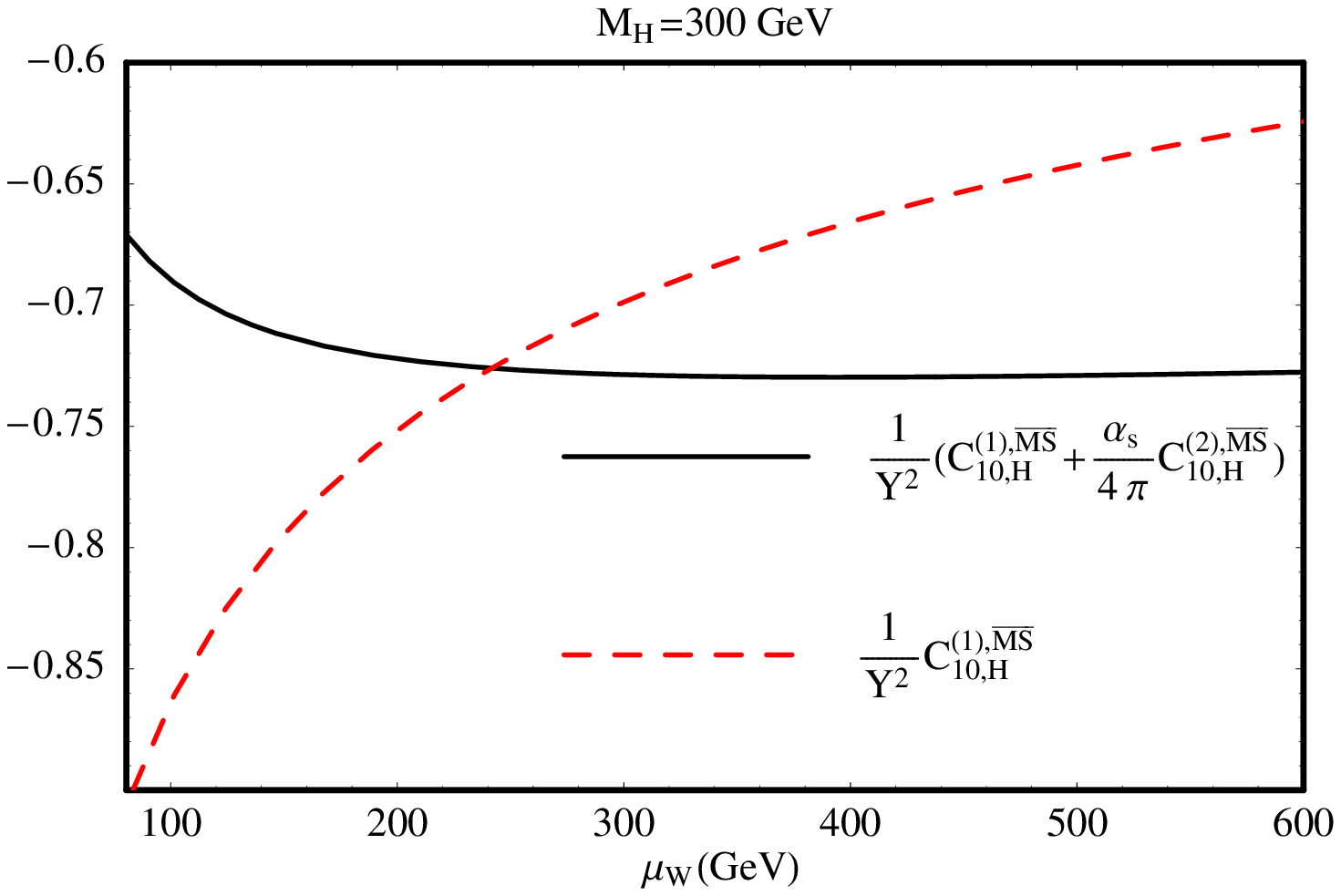,scale=0.8}
\end{center}
\caption{Dependence of the rescaled Wilson coefficient 
$\hat{C}_{10,\smallthdm}(\mu_b)$ on the matching scale $\mu_W$
(see Eq. (\ref{rescaled})) for 
$M_H=300$ GeV.
The dashed line shows the one-loop contribution expressed in
\ms-scheme for the $t$-quark mass, 
while the solid line includes the two-loop corrections in the same
scheme.}
\label{combimuw}
\end{figure}

Looking at the renormalization group equation (RGE) \cite{Gambino:2003zm} 
for $\hat{C}_{10,\smallthdm}$, one finds that $\hat{C}_{10,\smallthdm}$
does not run in QCD, i.e. 
\begin{equation}
\hat{C}_{10,\smallthdm}(\mu_b)=\hat{C}_{10,\smallthdm}(\mu_W) \, ,
\end{equation}
where the low scale $\mu_b$ is of the order of $m_b$.  
In Fig. \ref{combimuw} we show the dependence of 
$\hat{C}_{10,\smallthdm}(\mu_b)$ on the matching scale $\mu_W$ 
for $M_H=300$ GeV. It can be clearly seen that the inclusion 
of the two-loop contributions significantly lowers the dependence
on $\mu_W$. 
For $\mu_W >250$ GeV, $\hat{C}_{10,\smallthdm}(\mu_b)$ 
at two-loop precision is nearly $\mu_W$-independent. For $\mu_W$  
between $M_W$ and 250 GeV the two-loop Wilson coefficient varies  
about $\pm 4\%
$, whereas the corresponding 
one-loop coefficient varies about $\pm 11 \%
$. 


\section{Summary}\label{sec:summary}
The rare decay $\BXsll$ is subject of many contemporary analyses in particle physics since it is a promising channel in the search for
new physics beyond the SM. We computed NLO QED corrections to the matrix elements of effective operators and found that these corrections
include terms that are enhanced by large electromagnetic logarithms $\ln(m_b^2/m_\ell^2)$ whose numerical impact on the low-$q^2$
branching ratio is in the range of several percent. The zero of the forward backward asymmetry is also expected to get shifted by this
type of corrections.

We also showed that the inclusion of QCD corrections to the charged Higgs induced contributions to $C_{10}(\mu_W)$ in type-I and type-II
THDM significantly lowers the dependence of this Wilson coefficient on the scale $\mu_W$.

\section*{Acknowledgments}
We would like to thank Daniel Wyler, Enrico Lunghi, Thomas Gehrmann, and Miko{\l}aj Misiak for help during the preparation of our talks
and for a careful reading of this manuscript. We would also like to thank the organizers of the 40th conference ``Les Rencontres de Moriond
on QCD and High Energy Hadronic Interactions'' for a wonderful week in La Thuile. We also acknowledge support from the Marie Curie European
Grant. This work was supported by the Schweizerischer Nationalfonds.

\end{document}